\documentclass{iaus}
\usepackage{graphicx}

\title[The search for PopIII stars] 
{The search for Population III stars}

\author[S. di Serego Alighieri et al.]   
{Sperello di Serego Alighieri$^1$,
Jaron Kurk$^2$, Benedetta Ciardi$^3$, Andrea Cimatti$^4$, Emanuele
Daddi$^5$ \and Andrea Ferrara$^6$}

\affiliation{$^1$INAF -- Osservatorio Astrofisico di Arcetri, Largo E. Fermi 5,
Firenze, Italy \\ email: {\tt sperello@arcetri.astro.it} \\[\affilskip]
$^2$Max Planck Institut f\"ur Astronomie. K\"onigstuhl 17, Heidelberg, Germany
\\[\affilskip]
$^3$Max Planck Institut f\"ur Astrophysik, Karl Schwarzschild Str. 1,
Garching, Germany \\[\affilskip]
$^4$Universit\`a di Bologna, Via Ranzani 1, Bologna, Italy \\[\affilskip]
$^5$CEA/Saclay, Gif sur Yvette, France \\[\affilskip]
$^6$SISSA, Via Beirut 4, Trieste, Italy}

\pubyear{2008}
\volume{255}  
\pagerange{1--5}
\jname{Low-Metallicity Star Formation: From the First Stars to Dwarf Galaxies}
\editors{L.K. Hunt, S. Madden \& R. Schneider, eds.}
\begin{document}

\maketitle

\begin{abstract}
Population III stars, the first generation of stars formed
from primordial Big Bang material with a top--heavy IMF, 
should contribute substantially to the Universe reionization and they are
crucial for understanding the early metal enrichment of galaxies. Therefore
it is very important that these objects, foreseen by theories, are detected
by observations. However
PopIII stars, searched through the HeII 1640\AA\ line signature, have
remained elusive. We report about the search for the HeII line in a galaxy
at z=6.5, which is a very promising candidate. Unfortunately we are not yet
able to show the results of this search. However we call attention to the
possible detection of PopIII stars in a lensed HII dwarf galaxy at z=3.4,
which appeared in the literature some years ago, but has been overlooked.
\end{abstract}

\firstsection 
\section{Introduction}

Theoretical models foresee that the first generation of stars, forming from
primordial Big Bang material, should have unusually
massive stars (a {\it top--heavy} IMF), with masses up to about 500
$M_\odot$ (\cite[Schneider et al. 2002]{schn02}). These stars are called
Population III (PopIII) stars, and at least some of them are expected to
quickly release metals in the interstellar medium (ISM); therefore
later generations would soon be metal enriched. The metallicity threshold
in the ISM for producing PopIII stars should be around $10^{-5}$--$10^{-4}
Z_{\odot}$. The very massive PopIII stars would then produce a short phase
of unusually hard and strong UV radiation ($T_{eff} \sim 100000 K$),
resulting in a specific line emission signature (\cite[Shaerer
2002]{sch02}). The most prominent and unique emission line is expected to
be HeII 1640\AA, which can become as strong as 1/3 of Lyman $\alpha$,
making it observable with current 8--10m telescopes in the highest redshift
galaxies known (\cite[Scannapieco et al. 2003]{sca03}).

The detection of PopIII stars would be extremely important for
understanding the early chemical evolution of galaxies, and because they
should be crucial contributors to the reionization of the Universe
(\cite[Ciardi et al. 2003]{cia03}). PopIII stars have however remained
elusive: searches for HeII 1640\AA\ through stacking of spectra of Lyman $\alpha$
emitting galaxies (\cite[Dawson et al. 2004]{daw04} and \cite[Ouchi
et al. 2008]{ouc08}), through deep spectroscopy of an individual galaxy
(\cite[Nagao et al. 2005]{nag05}), or through dual (Ly$\alpha$ and HeII)
narrow--band imaging (\cite[Nagao et al. 2008]{nag08}) have failed.
\cite[Jimenez \& Haiman (2006)]{jim06} have ascribed to PopIII stars the HeII
1640\AA\ line detected in the composite spectrum of $\sim$1000 Lyman break 
galaxies at $z\sim 3$, but the line is only about 1/10 of Lyman $\alpha$ and is
considerably broader: it could therefore be attributed to a stellar wind 
feature associated with massive WR stars (\cite[Shapley et al. 2003]{sha03}). 
The upper limits to
the star formation rate (SFR) for PopIII stars, which can be set from these
negative results are now quite close to the rates expected from the models
of \cite[Tornatore et al. (2007)]{tor07} (see the contribution by T. Nagao to
these Proceedings).

We report about the search for the HeII 1640\AA\ line in a galaxy at z=6.5
and about the possible detection of PopIII stars in a lensed dwarf HII
galaxy at z=3.4, which has appeared in the literarure (\cite[Fosbury et al.
2003]{fos03}), but has so far been overlooked.

\section{The search for PopIII stars in KCS 1166}

KCS 1166 is a Lyman $\alpha$ emitting galaxy at z=6.518, discovered with
slitless spectroscopy in the atmospheric window at $\lambda\sim 9100$\AA\
(\cite[Kurk et al. 2004]{kur04}). In this object the Lyman $\alpha$ line is
clearly asymmetric, being steeper on the blue side, has a very large
equivalent width, at least 80\AA\ in the rest frame, and a luminosity of
$1.1\times 10^{43} erg\ s^{-1}$. No continuum is detected at shorter
wavelengths, while it is present redward of the line. These characteristics
and the fact that the HeII 1640\AA\ line is expected in a region of the
J--band relatively free of sky emissions make KCS 1166 a very good candidate for
the search of PopIII stars.

\begin{figure}[b]
\begin{center}
\includegraphics[width=13cm]{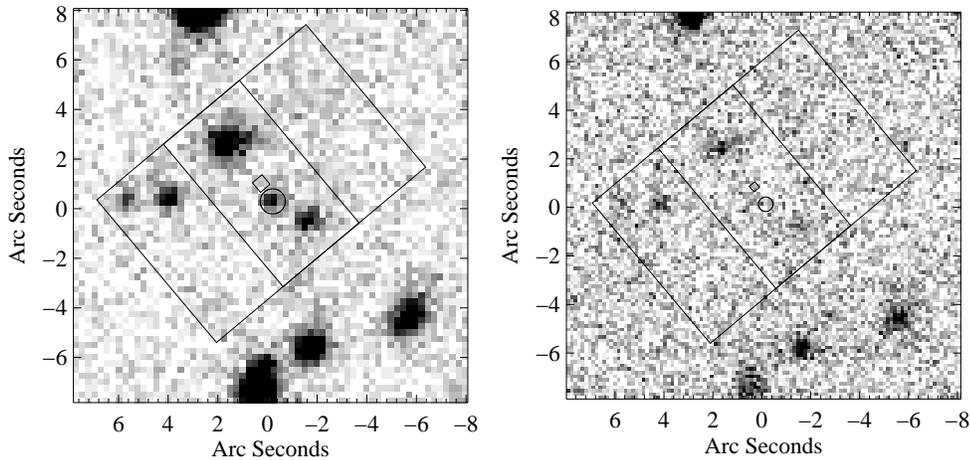} 
\caption{FORS z--band (left) and ISAAC J--band (right) images of the field of 
KCS 1166. The large boxes mark the two squared areas covered by the Integral
Field of view of SINFONI in  
the two observing positions, shifted by half of the field size to improve the sky
subtraction and to leave KCS 1166 in the overlapping area observed for the
whole observing time. The galaxy is marked by a small circle and is clearly visible
only in the z--band, which contains the Lyman $\alpha$ line.}

\label{fig1}
\end{center}
\end{figure}

We have therefore carried out near IR spectroscopy in the J--band of KCS
1166, using SINFONI, an Integral Field Spectrograph at the VLT 
(\cite[Bonnet et al. 2004]{bon04}), with a total on--source exposure time of
9 hours, equally spread over 3 consecutive nigths (Fig.\,\ref{fig1}).
Thanks to the excellent efficiency in the J--band and the lack
of OH Suppressor and of slit losses of SINFONI, we estimate
that our observations of KCS 1166 should a factor of about 1.5 more
efficient in detecting the eventual presence of the HeII 1640\AA\ line than
those of SDF J132440.6+273607 at z=6.3 by \cite[Nagao et al. (2005)]{nag05}, 
although we had a slightly shorter exposure time.

However at the time of this Conference we are yet unable to report about
the results of our observations, since the data reduction, which is rather
complicated for an Integral Field Spectrograph like SINFONI, is not yet finished.

\section{The possible detection of PopIII stars in the Lynx arc}

We take this opportunity to draw the attention of the Conference
participants and of those interested in PopIII stars to the results
obtained by \cite[Fosbury et al. (2003)]{fos03} on the Lynx arc, since
these could well mark the first detection of PopIII stars, and are completely
overlooked by the relevant literature. Although we recommend the interested
reader to read the paper by Fosbury et al. directly, we give here a summary of 
their work.

The Lynx arc is a dwarf HII galaxy at z=3.357, lensed by
the z=0.570 cluster RX J0848+4456, and it has been discovered serendipitously by
\cite[Holden et al. (2001)]{hol01} during spectroscopic follow--up of the
cluster. The HII galaxy is magnified by a factor of about 10, as evaluated
by a detailed analysis of the complex cluster environment from HST WFPC2
images. The unusual emission line spectrum has been thoroughly studied with the
LRIS, ESI and NIRSPEC at the Keck telescopes, covering the
rest--frame ranges 900--2500 and 3300--5700\AA. It
shows strong and self--absorbed Lyman $\alpha$ line and CIV 1548,1551\AA\
doublet, strong intercombination lines of NIV] 1483,1487\AA, CIII]
1907,1909\AA, and OIII] 1661,1666\AA, moderate HeII 1640\AA, absent [OII]
3727,3729\AA. The doublet SiIII] 1883,1892\AA\ is clearly detected in the
ESI spectrum, and is 40 times brighter, relative to H$\beta$, than foreseen
by models with scaled solar abundances. The intercombination lines have a
very narrow width ($\sigma \sim 30-35 km s^{-1}$), indicating a small
gravitating mass of about $10^9 M_{\odot}$, typical of a dwarf galaxy. 
The absence of NV 1240\AA\ and
the weakness of NIII] 1750\AA\ indicate that the ionizing source is a
blackbody, rather than a power law, as it would be expected in case of
ionization by an AGN.

The intensity of the continuum observed longward of Lyman $\alpha$ is completely explained
by the nebular continuum, as accurately predicted from the observed strength
of H$\beta$. The continuum produced by an instantaneous burst of $10^7
M_{\odot}$, Salpeter IMF, Z=0.05$\times Z_{\odot}$, age of 1 to 5 Myr would
be seen in the data, if it were present, but would produce an ionizing flux 20
times lower than the one necessary to produce the observed emission lines: this type 
of continuum, therefore, cannot be the source of ionization.

The photoionization model reproducing all the observed features has a
black--body ionizing source with: a) a black--body temperature of 80000K,
much higher than the effective temperature of Galactic compact HII regions,
which does not exceed 40000K; b) an ionization parameter U=0.1, also higher
than in local HII regions; c) a low {\it nebular} metallicity Z=0.05$\times
Z_{\odot}$. The lack of stellar continuum longward of Lyman $\alpha$, the
strong necessary ionizing flux, and the small stellar mass all indicate
that the ionizing flux should be produced by fewer than $10^6$ hot stars,
formed with a top--heavy IMF, and most likely with a metallicity still lower 
than the nebular one. The substantial overabundance of silicon in
the nebula indicate enrichment by past pair instability supernovae, as
those resulting from the total disruption of stars with 140--260
$M_{\odot}$ (\cite[Hegel \& Woosley 2002]{heg02}).

All these characteristics point strongly to the presence of PopIII stars,
some of which might have already exploded to partially enrich the ISM. However in
a dwarf galaxy like the Linx arc, the formation of PopII stars could have 
been delayed long enough to
make the uncontaminated UV signature of PopIII stars detectable at
intermediate redshift. In fact this uncontaminated signature of PopIII
stars might have a much shorter duration in the massive galaxies, which are
the only observable galaxies at high redshift. This might be the reason why
PopIII stars have not yet been detected at high redshift, where they have
been searched so far, following the obvious paradigm that primordial
material is more abundant in the very young Universe.

In a contribution to a conference \cite[Schaerer (2004)]{sch04} finds it
unlikely that the Lynx arc contains an extremely metal poor cluster,
because the ISM metallicity is 1/20 solar and there are no know cases
of the stellar metallicity lower than the nebular one, and the
alternative explanation of an obscured AGN (\cite[Binette et al.
2003]{Bin03}) seems to be capable of reproducing the observed spectrum.
However, it should not be surprising that new phenomena are observed when
dealing with a new class of objects, like PopIII stars, and Binette
et al. (2003) find it plausible that the stellar metallicity might be
lower than the nebular one. Furthermore, of the 5 models presented by
Binette et al. (2003) the one that best fits the observed line
ratios of the Lynx arc is the hot star (80000 K) model, which is relevant
for PopIII and which has only one inconsistent line, Si III] 1883,1892\AA,
whose discrepancy could in any case be due to the nucleosynthetic signature
of pair instability supernovae, as explained above. The other 4 models
also have discrepant Si III]
1883,1892\AA, but in addition have at least two more lines
inconsistent with the observations. Therefore the most likely explanations
for the observed spectrum of the Lynx arc remains the very hot star model
expected for PopIII stars.

\section{Final remarks}

Although we are yet unable to report on the results of our observations of
the HeII 1640\AA\ line in KCS 1166, which is a good candidate for PopIII
search at z=6.518, it is well possible that PopIII stars have already been
detected by \cite[Fosbury et al. (2003)]{fos03} in a dwarf star--forming
galaxy at z=3.357, thanks to the fortunate combination of the longer 
uncontaminated PopIII phase in dwarf galaxies and of the opportunity given
by the lensing magnification to identify PopIII stars, even if not at their
peak activity.

Clearly further detections of PopIII stars, or stringent upper limits,
would be extremely important to understand the connection between
reionization and metal production, since PopIII stars should be major players
on both scenes, and could spoil the simple proportionality between ionizing
photons and metals, claimed by some (see e.g. the contribution by A. Ferrara to
these Proceedings).

\section{Acknowledgements}

We would like to thank Bob Fosbury and Raffaella Schneider for very useful
comments.

\end{document}